\documentclass[twocolumn,prb,amsmath,amssymb,floatfix]{revtex4-2}
\usepackage{amsmath}
\usepackage{amssymb}
\usepackage{graphicx}
\usepackage{bm}
\usepackage{color}
\usepackage{hyperref}
\usepackage{graphicx}
\usepackage{upgreek}
\usepackage{scalerel}
\usepackage{multirow}
\usepackage{pgffor}
\usepackage{pdfpages}
\usepackage{mathdots}
\usepackage{float}
\usepackage{silence}
\usepackage{physics} 
\usepackage{lscape}   
\usepackage{color, colortbl}
\usepackage[table]{xcolor}
\usepackage{comment}  
\makeatletter
\AtBeginDocument{\let\LS@rot\@undefined}
\makeatother

\begin{document}

\title{Exceptional odd-frequency pairing  in non-Hermitian superconducting systems}
\author{Jorge Cayao}
\affiliation{Department of Physics and Astronomy, Uppsala University, Box 516, S-751 20 Uppsala, Sweden}
\author{Annica M. Black-Schaffer}
\affiliation{Department of Physics and Astronomy, Uppsala University, Box 516, S-751 20 Uppsala, Sweden}

\date{\today}
\begin{abstract}
We first show the realization of exceptional points in a non-Hermitian superconducting system based on a conventional superconductor and then demonstrate that, surprisingly, the system hosts odd-frequency pairing, solely generated by the non-Hermiticity. While there is a coexistence of even- and odd-frequency pairs under general conditions, we find that the even-frequency term vanishes at the exceptional degeneracies, leaving only odd-frequency pairing. This exceptional odd-frequency pairing is directly given by the imaginary part of the eigenvalues at the exceptional points and can be measured from the spectral function. Our results thus put forward non-Hermitian systems as a powerful platform to realize odd-frequency superconducting pairing.

\end{abstract}
\maketitle

\section{Introduction}
Superconductivity is a rare manifestation of quantum mechanics on a truly macroscopic scale and is also a basic ingredient in  emerging  quantum technologies \cite{Ac_n_2018}. To date, many  superconducting states have been reported, both intrinsic and engineered using conventional $s$-wave superconductors in proximity to other materials, such as topological superconductivity in various hybrid devices \cite{frolov2020topological,flensberg2021engineered}. While the scheme for creating unconventional superconductors may differ,  their properties are always to a very large extent dictated by the symmetries of their fundamental constituents, the electron, or Cooper, pairs.  

The Cooper pair wavefunction, or \textit{pair amplitude}, depends on the degrees of freedom of the paired electrons \cite{Tinkham}. While all the degrees of freedom are important  for the Cooper pair symmetries,  it is perhaps the time  at which electrons pair  that introduces the most interesting but least explored properties, mainly due to their relevance  in dynamic quantum matter \cite{DynamicMatter}.  In its most general form,   electrons can pair at different times, or equivalently at finite frequency $\omega$. This enables \textit{odd-frequency} (odd-$\omega$) pairing, where the pair amplitude is odd in  relative time, or equivalently odd in $\omega$. Odd-$\omega$ pairing is thus an intrinsically dynamic and time-dependent effect \cite{RevModPhys.77.1321,Nagaosa12,Balatsky2017,cayao2019odd,triola2020role}. 

Since its initial conception \cite{bere74}, odd-$\omega$ pairing has generated an ever increasing interest, not only due to its  dynamical nature but also because it explains several exotic effects, such as long-range proximity effects or paramagnetic Meissner signatures \cite{RevModPhys.77.1321,Nagaosa12,Balatsky2017,cayao2019odd,triola2020role}. Interestingly, odd-$\omega$ pairs have been shown to emerge in several systems using just conventional $s$-wave superconductors, with notable examples  in superconducting heterostructures \cite{PhysRevLett.86.4096,Kadigrobov01,PhysRevB.76.054522,PhysRevB.87.220506,PhysRevB.92.100507,PhysRevB.96.155426,PhysRevB.97.134523}, multiband superconductors \cite{PhysRevB.88.104514,PhysRevB.90.220501,PhysRevB.92.094517,PhysRevB.93.201402,10.1093/ptep/ptw094,Eschrig2007}, and  time-periodic superconductors \cite{PhysRevB.94.094518,PhysRevB.103.104505}. Still, these systems share a common characteristic in that all  represent  closed systems, described by Hermitian Hamiltonians.

Physical systems are, however, always coupled to their environment, and thus open, 
where dissipative effects are unavoidable and described by non-Hermitian (NH)  processes \cite{Moiseyev}. Notably, dissipation has  been shown to lead unique NH effects that broadens the system symmetries \cite{PhysRevX.9.041015}, giving rise to unusual phases \cite{el2018non,RevModPhys.93.015005,doi:10.1080/00018732.2021.1876991} with no analog in Hermitian setups.  The main property of NH systems is that they exhibit a complex spectrum with level degeneracies, known as \textit{exceptional points} (EPs) \cite{TKato,heiss2004exceptional,berry2004physics,Heiss_2012,PhysRevLett.86.787,PhysRevLett.103.134101,PhysRevLett.104.153601,gao2015observation,doppler2016dynamically}, where eigenstates and eigenvalues coalesce, in stark contrast to Hermitian systems.
Moreover, non-Hermiticity not only allows  to understand and engineer dissipative systems, but it can also be precisely controlled and hence used for sophisticated applications \cite{el2018non,RevModPhys.93.015005,doi:10.1080/00018732.2021.1876991},  such as for high-performance lasers \cite{feng2014single,peng2016chiral,hokmabadi2019non,parto2020non} and sensors \cite{chen2017exceptional,hodaei2017enhanced,wiersig2020review,PhysRevLett.125.180403}.

Non-Hermiticity has also recently been shown to ramify the particle-hole symmetry \cite{PhysRevX.9.041015},  intrinsic in superconductors. It is thus natural to ask about its impact on the symmetry of the pair amplitude. Moreover,  due to the close link between non-Hermiticity and dissipation, which reflects a dynamical essence, it represents a genuinely promising ground to explore as origin of odd-$\omega$ pairing. However, the connection between non-Hermiticity and odd-$\omega$ pairing has so far received little attention, with
studies only focusing on symmetry classification \cite{bandyopadhyay2020classification} or spectral broadening in a Dynes superconductor \cite{PhysRevB.102.014508}. 
This has left, for example, the role of the main NH characteristic, the EPs, completely unexplored.

\begin{figure}[!t]
\centering
	\includegraphics[width=0.5\columnwidth]{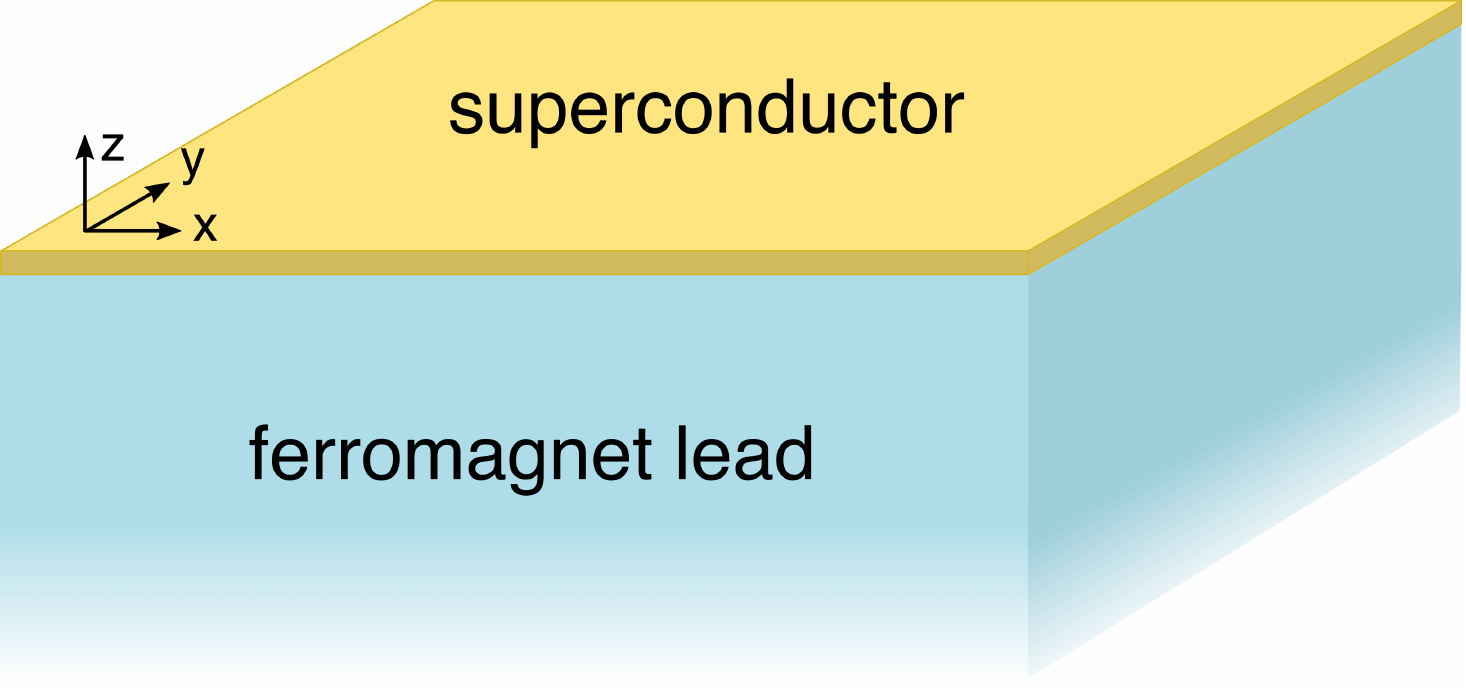}
	\caption{Sketch of a  2D conventional  $s$-wave superconductor coupled to a semi-infinite ferromagnet lead.  Due to coupling to the lead, the total system is described by an effective non-Hermitian Hamiltonian.}
\label{Fig1} 
\end{figure}

In this work we first show how NH superconducting systems easily host odd-$\omega$ pairing, entirely due to non-Hermiticity. Surprisingly, we find that all even-$\omega$ pairing vanishes at the EPs, leaving only a large odd-$\omega$ contribution, which we refer to as {\it exceptional odd-$\omega$ pairing}. 
We then illustrate these results in a realistic NH system consisting of a conventional superconductor coupled to a ferromagnet lead, see Fig.\,\ref{Fig1}. 
Finally, we show that the exceptional odd-$\omega$ pairing, as well as the EPs, can be detected in the spectral function via angle-resolved photoemission spectroscopy (ARPES). Our findings thus  put forward an entirely different route for generating odd-$\omega$  pairing, paving the way for NH engineering of dynamical superconducting states.

 \section{Pair amplitudes}
To understand how odd-$\omega$ pairing appears in NH systems,  we first inspect the  structure of the  pair amplitude $F$, which is obtained from the electron-hole (eh), or anomalous, part of the Green's function $G(\omega)=(\omega-H)^{-1}$ \cite{mahan2013many,zagoskin}. Here $H$ is the system Hamiltonian in Nambu space  $\psi=(c,c^{\dagger})^{\rm T}$, where $c$ annihilates an electronic state. While $F$ can be directly found from a matrix inversion, to gain basic understanding of its dependencies it is more useful to   express  $(\omega-H)^{-1}$ in terms of its adjugate ({\rm Adj}) and determinant ({\rm det})  \cite{HornRoger}. In this way, $F$ reads
\begin{equation}
\label{EQ1}
F(\omega)=\frac{1}{{\rm det}(\omega-H)}[{\rm Adj}(\omega-H)]_{\rm eh}\,,
\end{equation}
 with ${\rm Adj}(.)$   found as the transpose of the cofactor matrix \cite{HornRoger}. The representation of $F$ in Eq.\,(\ref{EQ1}) is general and valid for  both Hermitian and NH Hamiltonians.   

While Eq.\,(\ref{EQ1}) might  seem complicated, it actually offers a simple way to analyze how odd-$\omega$ pairing appears, as any odd-$\omega$ part must come either from the denominator or numerator.
For this reason, we first note that the poles of $G$ give the quasiparticle energies, or the eigenvalues $E_{i}$ of $H$. Then, to visualize the appearance of odd-$\omega$ pairing in Eq.\,(\ref{EQ1}) it is convenient  to express the determinant in terms of $E_{i}$: ${\rm det}(\omega-H)=\Pi_{i}(\omega-E_{i})$  \cite{HornRoger}.   For simplicity, but without loss of generality, we for now assume that spin, space, and orbital are not  active degrees of freedom, such that $H$ only has two eigenvalues $E_{1,2}$. Thus, we can write ${\rm det}(\omega-H)=(\omega-E_{1})(\omega-E_{2})$, with $E_{1,2}$  related by particle-hole symmetry, which can differ for Hermitian and NH Hamiltonians \cite{PhysRevX.9.041015}.

For Hermitian systems, $E_{1,2}=\pm E$ and the denominator of  Eq.\,(\ref{EQ1}) becomes  ${\rm det}(\omega-H)=\omega^{2}-E^{2}$, clearly an  even function of  $\omega$. Also, the numerator of Eq.\,(\ref{EQ1}), ${\rm Adj}(.)$, does not develop any odd-$\omega$ term in this simple case.  However, we have verified that in systems with finite odd-$\omega$ pairing, such as two-band superconductors \cite{triola2020role}, it is the $[{\rm Adj}(.)]_{\rm eh}$ term that generates odd-$\omega$ pairing, while ${\rm det}(.)$  only  provides even powers of $\omega$. Thus, for  time-independent Hermitian Hamiltonians with the properties discussed above, the only option for $F$ to contain odd-$\omega$ pairing comes from the $[{\rm Adj}(.)]_{\rm eh}$ matrix.

In contrast, for NH systems the  eigenvalues are no longer real (Re) but develop an imaginary (Im) term, $E_{n}=a_{n}-i b_{n}$, with $a, b$  both real-valued numbers \footnote{Note that, without loss of generality, we consider negative imaginary terms of the eigenvalues but similar conclusions are obtained with positive imaginary terms. Moreover, here we do not consider PT-symmetric systems which can exhibit real spectra despite being non-Hermitian, see e.g.~\cite{doi:10.1080/00018732.2021.1876991}.}. For NH superconducting systems, they come in pairs, obeying $E_{1}=-E_{2}^{*}$ due to the charge-conjugation symmetry  \cite{pikulin2012topological,pikulin2013two,san2016majorana,avila2019non,PhysRevX.9.041015}. This imposes $a_{1}=-a_{2}=a$ and $b_{1}=b_{2}\equiv b$. Then,   the denominator  in Eq.\,(\ref{EQ1})  reads ${\rm det}(\omega-H)=\omega^{2}-a^{2}-b^{2}+2i\omega b$, where the last term now directly reveals an odd-$\omega$ term proportional to $b$, while the numerator of Eq.\,(\ref{EQ1}) still does not contain any odd-$\omega$ part.  Taken together, the pair amplitude of NH systems reads
\begin{equation}
\label{EQ2}
F_{\rm NH}(\omega)=\frac{[{\rm Adj}(\omega-H)]_{\rm eh}}{d^{2}+4\omega^{2}b^{2}}(d-2i\omega b)\,,
\end{equation}
where $d=\omega^{2}-a^{2}-b^{2}$ is an even function of $\omega$. This $F_{\rm NH}$  has  both even- and odd-$\omega$ parts, proportional to $d$ and $i\omega b$, respectively. Importantly, the odd-$\omega$ term is purely driven by the Im part of the eigenvalues, $b$.   

The main characteristic  of NH Hamiltonians is the presence of EPs, where eigenvalues and eigenvectors coalesce \cite{TKato,heiss2004exceptional,berry2004physics,Heiss_2012}. This implies that at the EPs, $a_{1}=-a_{2}=0$ and $b_{1}=b_{2}=b$, leaving  a single purely Im eigenvalue, $E_{1,2}=ib$. Also, then  $d=\omega^{2}-b^{2}$, which vanishes when $\omega=|b|$, i.e.~at the EP. Hence, at the EP, the even-$\omega$  term of $F_{\rm NH}$  vanishes,  leaving only  odd-$\omega$ pairing, which we refer to as \textit{exceptional odd-$\omega$ pairing}. We thus conclude that  odd-$\omega$ pairing can  be easily induced in a NH system, even when it is completely absent in the Hermitian regime, and even more interestingly, it becomes the only  source of  pairing at  EPs.

\section{Realization of a NH superconducting system}
Next we show that odd-$\omega$ pairing emerges naturally in realistic NH systems. For this purpose, we first engineer a simple NH superconducting system by coupling a conventional spin-singlet $s$-wave 2D superconductor \cite{PhysRevB.93.155402,kjaergaard2016quantized,PhysRevLett.119.176805,bottcher2018superconducting,casparis2018superconducting,o2021epitaxial,lutchyn2018majorana,zhang2019next,prada2020andreev} to a ferromagnetic lead,  see Fig.\,\ref{Fig1}.   This NH system is modeled by the following effective Nambu Hamiltonian 
\begin{equation}
\label{modelNH}
H_{\rm eff}=H_{\rm S}+\Sigma^{r}(\omega=0)\,,
\end{equation}
where $H_{\rm S}=\xi_{k}\tau_{z}-\Delta \sigma_{y}\tau_{y}$ 
describes the (closed) superconductor in the basis  $(c_{k,\uparrow},c_{k,\downarrow},c_{-k,\uparrow}^{\dagger},c_{-k,\downarrow}^{\dagger})$, with $c_{k,\sigma}$ annihilating an electron with momentum $k$ and spin $\sigma$. Here
 $\xi_{k}=\hbar^{2}k^{2}/2m-\mu$ is the kinetic energy with $k=(k_{x},k_{y})$, $\sigma_{i}$ and $\tau_{i}$ the spin and electron-hole Pauli matrices, respectively, $\mu$ is the chemical potential, and $\Delta$ is the spin-singlet $s$-wave pair potential.  
We consider either intrinsic thin film superconductors or proximity-induced superconductivity into a thin film semiconductor, both effectively producing a 2D superconductor, but our results are also valid in the interface region for 3D superconductors 
 \footnote{As we do not consider in-plane inhomogeneities in the junction, the superconducting order parameter does not have to be determined in a self-consistent calculation, but $\Delta$ represents the effective order parameter in the superconductor, set by material specific parameters before coupling to the lead.}. Further, $\Sigma^{r}(\omega=0)$ is the retarded spin-dependent self-energy  at $\omega=0$ describing the effect of the lead on the superconductor. While $\Sigma^{r}$, in general, depends on $\omega$, its independence of $\omega$ is well justified  e.g. in the wide band limit \cite{datta1997electronic,kohler2005driven,Ryndyk2009,PhysRevResearch.1.012003}. With the lead being semi-infinite, $\Sigma^{r}$ has both  Re and  Im terms. While the Re part is Hermitian and just renormalizes the elements of $H_{\rm S}$, the Im part is NH and introduces dramatic changes, which becomes our focus here \cite{datta1997electronic,Ryndyk2009,PhysRevResearch.1.012003}. We obtain $\Sigma^{r}(\omega=0)={\rm diag}(\Sigma^{r}_{\rm e},\Sigma^{r}_{\rm h})$   analytically, see Supplemental Material (SM) for details \cite{SM}, where we  approximate \footnote{This approximation is valid e.g.~in the wide band limit commonly used in quantum transport \cite{datta1997electronic,Ryndyk2009}.} 
\begin{equation}
\label{Sigmaeh}
\Sigma^{r}_{\rm e,h}(\omega=0)=-i \Gamma \sigma_{0}-i\gamma \sigma_{z}\,,
\end{equation}
with $\Gamma=(\Gamma_{\uparrow}+\Gamma_{\downarrow})/2$ and $\gamma=(\Gamma_{\uparrow}-\Gamma_{\downarrow})/2$. Here, $\Gamma_{\sigma}=\pi|t'|^{2}\rho_{\rm L}^{\sigma}$ with  $\rho_{\rm L}^{\sigma}$  the surface  density of states of the lead (L) for spin $\sigma=\uparrow, \downarrow$, controlled by the Zeeman field in the ferromagnet, and $t'$  the hopping amplitude into the lead from the superconductor.  For obvious reasons we refer to $\Gamma_{i}$ as to the coupling amplitude.   Due to causality, all terms in $\Sigma^{r}$ reside in the lower complex energy half-plane, a clear signal of dissipation.

\begin{figure}[!t]
\centering
	\includegraphics[width=0.99\columnwidth]{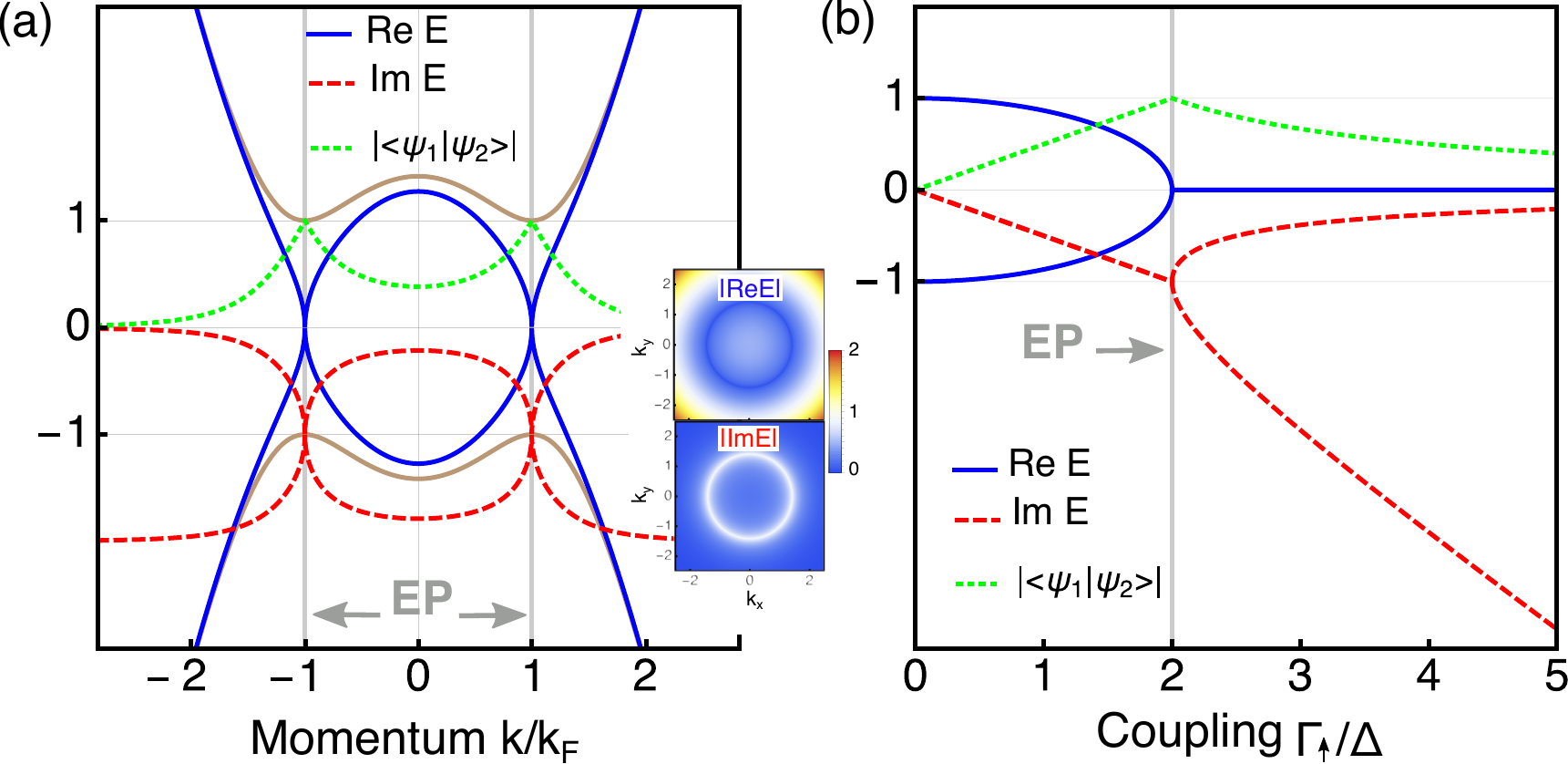}
	\caption{Re (blue) and Im (red) parts of the eigenvalues in Eq.~\eqref{eigenvaluesEP} as a function of $k$ at fixed $\Gamma_{\uparrow,\downarrow}$ (a) and as a function of $\Gamma_{\uparrow}$ at fixed $\xi_{k}=0$ and  $\Gamma_{\downarrow}=0$ (b), with wavefunction overlap in green. At the EP transition (gray) the eigenvalues coalesce and the wavefunctions become parallel.  Brown curve shows eigenvalues without non-Hermiticity. Inset depicts the absolute value of Re and Im parts of the eigenvalues. Parameters: $\Gamma_{\uparrow}=2$, $\Gamma_{\downarrow}=0$, $\Delta=1$\, $\mu=1$, $k_{\rm F}=\sqrt{2m\mu/\hbar^{2}}$.}
\label{Fig2} 
\end{figure}

Using Eq.~\eqref{Sigmaeh}, the  eigenvalues of $H_{\rm eff}$ are given by
\begin{equation}
\label{eigenvaluesEP}
E_{n}=-i\Gamma\pm\sqrt{\Delta^{2}+\xi_{k}^{2}-\gamma^{2}\pm 2i|\xi_{k}||\gamma|}\,,
\end{equation}
which acquire Im terms solely due to the effect of the lead through $\Gamma$ and $\gamma$. At  $\Gamma=\gamma=0$,  the system is Hermitian with real eigenvalues $E_{n}=\pm\sqrt{\Delta^{2}+\xi_{k}^{2}}$, shown in brown in Fig.\,\ref{Fig2}(a).
At any non-zero coupling, $E_{n}$ develops  non-zero Im terms, a clear feature of NH physics.  The inverse  of ${\rm Im} (E_{n})$ represents  the average time a quasiparticle remains in the superconductor before escaping into the lead, setting the length scale $\ell_{\Gamma}=\hbar v_{\rm F}/{\rm Im} (E_{n})$, with $v_{\rm F}$ the Fermi velocity in the superconductor, for how deep the NH effect penetrates if using a 3D superconductor. At  $\Gamma_{\uparrow}=\Gamma_{\downarrow}$, $\gamma=0$ and all $E_{n}$'s acquire the same Im term, equal to $-i\Gamma$. It is only when $\Gamma_{\uparrow}\neq\Gamma_{\downarrow}$ that all $E_{n}$'s undergo the special transition at which their Re and Im parts merge into a single value, $i\Gamma$, thus producing  EPs. This occurs when the square root in Eq.\,(\ref{eigenvaluesEP}) vanishes
\begin{equation}
\label{EPConditions}
\Delta^{2}+\xi_{k}^{2}-\gamma^{2}=0\, \quad \text{and} \quad   2i|\xi_{k}||\gamma|=0 \,.
\end{equation}
To visualize these EP conditions, we present in Fig.\,\ref{Fig2}(a,b) the Re (solid blue) and Im (dashed red) parts of $E_{n}$  as a function of $k$ and $\Gamma_{\uparrow}$, with the EP transitions marked in gray.  We observe that the electron- and hole-like $E_{n}$ coalesce, and EPs appear, only at $\xi_{k}=0$, or equivalently $k=\sqrt{2m\mu/\hbar^{2}}$, provided $\Delta=|\gamma| \neq 0$. The EPs extend into a circle when $k$ is plotted in 2D, see inset in Fig.\,\ref{Fig2}. As expected for EPs, the conditions in Eqs.\,(\ref{EPConditions}) not only define the coalescence of $E_{n}$, but they also define the coalescence of the associated eigenvectors. In fact, at the EPs, the associated wavevectors become parallel instead of orthogonal as for Hermitian systems, as seen by  their scalar product (dotted green) in Fig.\,\ref{Fig2}. In Fig.\,\ref{Fig2}(b), we instead fix $\xi_{k}=0$ and plot the eigenvalues as a function of $\Gamma_{\uparrow}$ at fixed $\Gamma_{\downarrow}=0$ and again see a clear EP transition. 
Thus, our simple, but physical, NH superconducting system in Fig.~\ref{Fig1} host clear and stable EPs,  which represent  the main  property of NH systems \cite{RevModPhys.93.015005,doi:10.1080/00018732.2021.1876991}.

\section{Exceptional odd-$\omega$ pair amplitude}
Having established the existence of EPs in the NH system in Fig.~\ref{Fig1} and Eq.~\eqref{modelNH}, we next turn to calculating its pair amplitudes using the anomalous components of the  retarded Green's function $G^{r}=(\omega-H_{\rm eff})^{-1}$. We obtain even- and odd-$\omega$ (E,O) pair amplitudes given by 
\begin{equation}
\label{FNH}
\begin{split}
F_{\uparrow\downarrow}^{\rm E}(\omega)=\frac{-\Delta Q_{\uparrow\downarrow}}{Q_{\uparrow\downarrow}^{2}+4\omega^{2}\Gamma^{2}}\,,\quad
F_{\uparrow\downarrow}^{\rm O}(\omega)=\frac{-2i\omega\Delta \Gamma}{Q_{\uparrow\downarrow}^{2}+4\omega^{2}\Gamma^{2}}\,,
\end{split}
\end{equation}
where $Q_{\uparrow\downarrow}=\Delta^{2}+\xi_{k}^{2}+\Gamma^{2}- \gamma^{2}-\omega^{2}-2i\gamma\xi_{k}$ is an even function in $\omega$. Likewise, we get  $F_{\downarrow\uparrow}^{\rm E (O)}=-F_{\uparrow\downarrow}^{\rm E(O)}(\Gamma_{\uparrow}\leftrightarrow \Gamma_{\downarrow})$, but we do not find any equal spin pairing. An interesting feature is that $F^{\rm O}_{\uparrow\downarrow}$ is proportional to $\Gamma$, showing that it is a direct   NH result, as in Eq.~\eqref{EQ2}.   The finite pair amplitudes can also be interpreted as a result of Andreev reflection at the superconductor-lead interface \cite{Pannetier2000,Klapwijk2004,Cayao_2017,Cayao_2018}.

\begin{figure}[!t]
\centering
	\includegraphics[width=0.99\columnwidth]{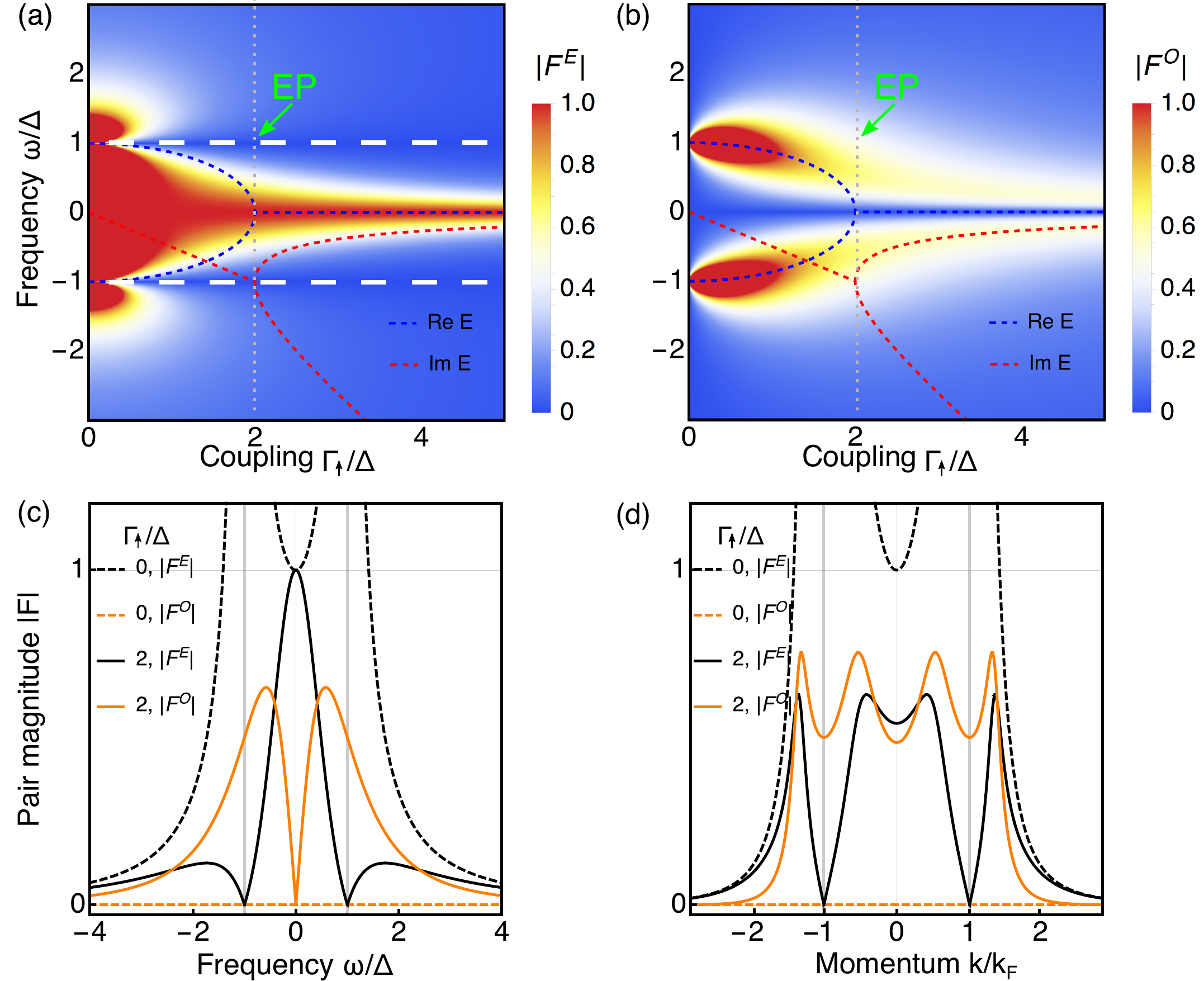}
	\caption{Absolute value of  even- (a) and odd-$\omega$ (b) pair amplitudes as a function of $\omega$ and $\Gamma_{\uparrow}$ at $\xi_{k}=0$, $\Gamma_{\downarrow}=0$, with  the color scale cut off at 1 for visualization. Dashed blue and red curves show the Re and Im parts of the eigenvalues, respectively. Also marked are the EP transition lines (grey) and energy values (green arrows), and values where the even-$\omega$ pairing vanish (dashed white). Pair amplitudes as a function of $\omega$ at   $\xi_{k}=0$ (c) and as a function of $k$ at fixed $\omega/\Delta=1$ (d) for different values of $\Gamma_{\uparrow}$. Rest of parameters are as in Fig.\,\ref{Fig2}.}
\label{Fig3} 
\end{figure}

To further inspect the NH effect on $F_{\downarrow\uparrow}^{\rm E, O}$, we plot their absolute values in Fig.\,\ref{Fig3} as a function of $\omega$, $\Gamma_{\uparrow}$, and  $k$.
At  $\Gamma=\gamma=0$, the system is Hermitian and then only the even-$\omega$ part survives, as seen both in Eqs.\,(\ref{FNH}) and Fig.\,\ref{Fig3}. 
At finite coupling, the system becomes NH and even- and odd-$\omega$ pairs generally coexist. As seen in Fig.\,\ref{Fig3}, both pair amplitudes  develop large values, but in different regimes, allowing us to establish a clear distinction between them: While $F^{\rm E}_{\uparrow\downarrow}$  is large around $\omega=0$,  $F^{\rm O}_{\uparrow\downarrow}$ exhibits surprisingly similarly large values at higher $\omega$ \footnote{While small to moderate values of dissipation, via  $\Gamma_{\uparrow,\downarrow}$, induce NH odd-$\omega$ pairs, we note that very large values can destroy superconductivity, as evident from Eq.~\eqref{FNH}. For obvious reasons we do not consider this latter regime}.

Next we examine the effect of EPs on   $F^{\rm O,E}_{\uparrow\downarrow}$ in Eqs.\,(\ref{FNH}). For this reason we analyze the term $Q_{\uparrow\downarrow}$ at the EPs, where the latter are defined by the conditions in Eqs.\,(\ref{EPConditions}) and  only present for $\gamma\neq0$. By using these EP conditions, we get $Q_{\uparrow\downarrow}=\Gamma^{2}-\omega^{2}$, assuming we already have tuned   $\Delta=|\gamma|$. Interestingly, $Q_{\uparrow\downarrow}$ vanishes exactly at $\omega=|\Gamma|$, i.e.~exactly at the magnitude of the eigenvalues at  EPs, see Eqs.\,(\ref{eigenvaluesEP}). 
Thus, at the EPs, we find only odd-$\omega$ pairing $F^{\rm O}_{\uparrow\downarrow}(\omega)=-(i\Delta)/(2\omega\Gamma)$, with $|\omega|=\Gamma$, as the even-$\omega$ part identically vanishes.
This exceptional odd-$\omega$ pairing is unusual for two additional reasons: its size is solely determined by the NH processes $\Gamma$ and $\gamma$, as $|\omega|=\Gamma$ and $\Delta = |\gamma|$ at the EPs, and it has a clear ${\rm sgn}(\omega)/\omega^{2}$ behavior,  unlike Hermitian systems \footnote{Although the ${\rm sgn}(\omega)/\omega^{2}$ behavior of the exceptional odd-$\omega$ pairing has not been found in other systems, odd-$\omega$ pairing acquires an interesting dependence in topological superconductors where it exhibits a $1/\omega$  behavior around $\omega=0$ in the presence of Majorana states \cite{cayao2019odd}.}. 
In Fig.\,\ref{Fig3}(a), the vanishing of the even-$\omega$ pairing actually occurs along the whole line  $\omega=\Delta$ as $\Gamma_{\uparrow}$ is varied (white dashed line), although the EP only occurs at the point $\Gamma_{\uparrow}/\Delta=2$ and at  $\omega/\Delta=1$ in this plot (green arrow). This is because 
the particular choice of parameters in Fig.\,\ref{Fig3}(a) results   in $Q_{\uparrow\downarrow}=0$ and thus zero even-$\omega$ pairing for all $|\omega|=\Delta$; note that the second condition for EPs, $\xi_{k}=0$, in Eqs.\,(\ref{EPConditions}), is  satisfied here. We thus find that vanishing even-$\omega$ pairing is intimately related to the occurrence of EPs in our system, leaving only finite exceptional odd-$\omega$ pairing, which, in turn, is solely determined by the magnitude of the eigenvalues at the EPs.

\section{Spectral signatures}
To detect the  EPs and  the odd-$\omega$ pairing,  we  study the spectral function $A(\omega,k)=-{\rm Im}{\rm Tr}(G^{\rm r}-G^{\rm a})$ \cite{mahan2013many,zagoskin} accessible via e.g.~ARPES measurements  \cite{hufner2013photoelectron,lv2019angle,yu2020relevance}, where $G^{\rm a}=[G^{\rm r}]^{\dagger}$ is the advanced Green's function \footnote{With the advent of high-resolution ($\sim70$\,$\mu$eV)    ARPES, also superconductors with low critical temperatures are accessible \cite{doi:10.7566/JPSJ.84.072001,RevModPhys.93.025006}, but alternative probes might also involve transport across the junction, such as conductance measurements \cite{kjaergaard2016quantized}.}. 
To elucidate the  pair amplitude dependency, it is useful to write the diagonal entries of $G^{\rm r}$ in terms of the  pair amplitudes. The diagonal electron terms are thus given by
\begin{equation}
[G_{0}^{r}(\omega)]_{\uparrow\uparrow(\downarrow\downarrow)}=\pm\frac{(\omega+\xi_{k}+i\Gamma_{\downarrow(\uparrow)})}{\Delta}[F(\omega)]_{\uparrow\downarrow(\downarrow\uparrow)}\,,
\end{equation}  
with $F_{\uparrow\downarrow}=F_{\uparrow\downarrow}^{\rm E}+F_{\uparrow\downarrow}^{\rm O}$  given by Eqs.\,(\ref{FNH}). 
The diagonal hole terms are $[\bar{G}_{0}^{r}]_{\uparrow\uparrow(\downarrow\downarrow)}=[G_{0}^{r}]_{\downarrow\downarrow(\uparrow\uparrow)}(\xi_{k}\rightarrow-\xi_{k},\Gamma_{\uparrow(\downarrow)}\rightarrow\Gamma_{\downarrow(\uparrow)})$. 
We  further isolate the individual even- and odd-$\omega$ pair  contributions by writing  $A=A^{\rm E}+A^{\rm O}$ with $A^{\rm E(O)}$ being due to $F_{ab}^{\rm E(O)}$. 
\begin{figure}[!t]
\centering
	\includegraphics[width=0.95\columnwidth]{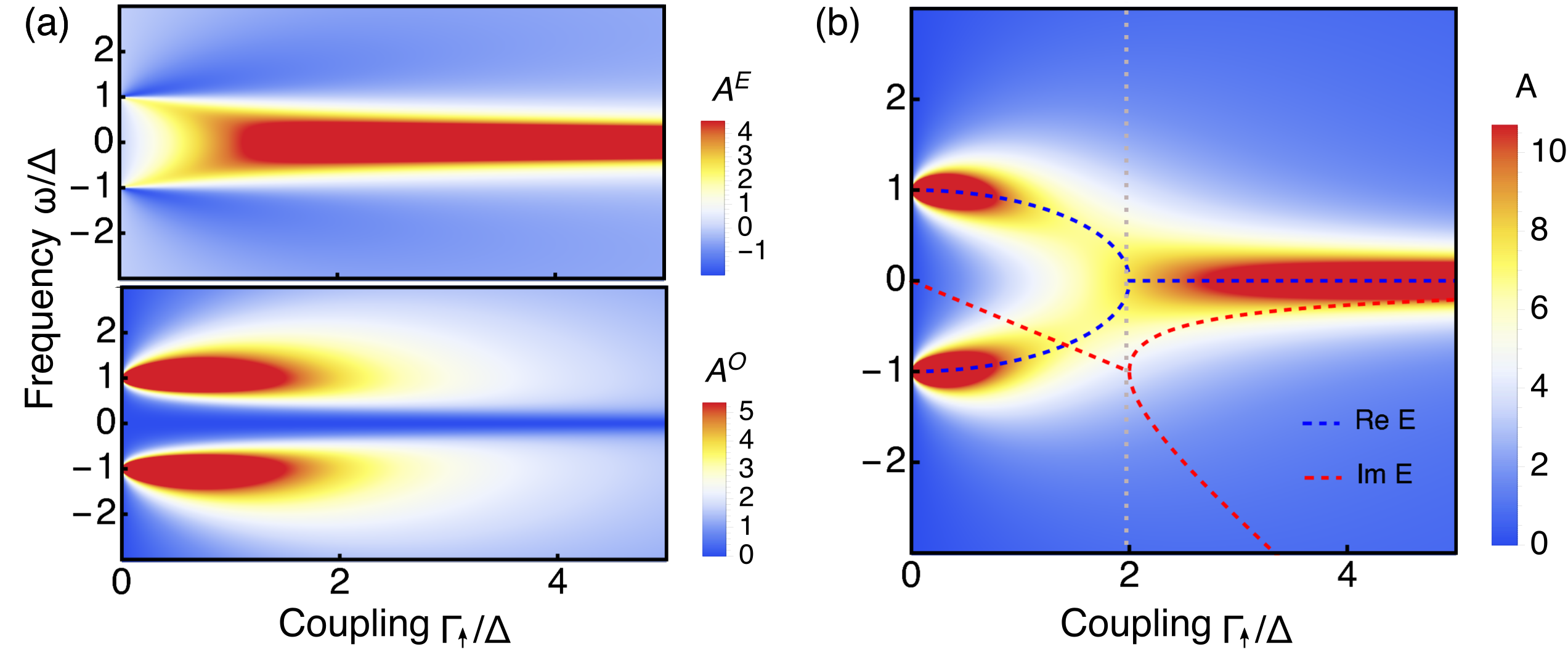}
	\caption{(a) Spectral function $A$ divided into individual contributions $A^E$ from even- (top) and $A^O$ from odd-$\omega$ (bottom) pairing as a function of $\omega$ and $\Gamma_{\uparrow}$ at $\xi_{k}=0$, $\Gamma_{\downarrow}=0$. (b) Total spectral function $A=A^{\rm E}+A^{\rm O}$, with EP transition lines (grey) and Re (dashed blue) and Im (dashed red) parts of eigenvalues depicted. Rest of parameters are as in Figs.\ref{Fig2}-\ref{Fig3}.}
\label{Fig4} 
\end{figure}

In  Fig.\,\ref{Fig4} we plot $A^{\rm E,O}$ and $A$ as functions of $\omega$ and $\Gamma_{\uparrow}$  at $\xi_{k}=0$.
By examining  the individual contributions in Fig.\,\ref{Fig4}(a), we note that they exhibit large values in different ranges of $\omega$ and $\Gamma_{\uparrow}$. In fact,  $A^{\rm E}$ acquires large values around $\omega=0$ and high $\Gamma_{\uparrow}$, similar to $F^{\rm E}_{\uparrow\downarrow}$ in Fig.\,\ref{Fig3}(a). Surprisingly, it also becomes negative for some parameters. On the other hand, $A^{\rm O}$ instead shows large values at finite $\omega$ and low $\Gamma_{\uparrow}$, stemming from large  $F^{\rm O}_{\uparrow\downarrow}$ for the same parameters, see Fig.\,\ref{Fig3}(b). The total  spectral function $A$  in Fig.\,\ref{Fig4}(b) captures the main features of both $A^{\rm E}$ and $A^{\rm O}$, where $A^{\rm O}$  also compensates for the negative values of $A^{\rm E}$. Note that $A$ also clearly signals the EP transition  (gray line). In fact, at the EP energy, $|\omega|=\Gamma$,  we estimate  $A = A^{\rm O} \approx 2\omega F^{\rm O}/\Delta$, with $F^{\rm O}=\Delta/(\omega\Gamma)$ being the magnitude of the exceptional odd-$\omega$ pairing. Thus, the spectral function detects the EP transition which then allows  to measure the  exceptional odd-$\omega$ pairing.

Experimentally, to generate exceptional odd-$\omega$ pairing, high control of $\Delta$ and $\Gamma_{\sigma}$ is necessary. For $\Delta$, recent works have reported well-controlled proximity-induced superconductivity in only $a=7$\,nm thick InAs films with $\Delta=0.2$\,meV and tunable using interface barriers \cite{PhysRevB.93.155402}. 
For $\Gamma_{\sigma}$, both the spin-dependent  density of states and the tunneling between lead and superconductor can be tuned, see Eq.\,(\ref{Sigmaeh}). Here, the Zeeman field of the lead guarantees distinct $\Gamma_{\sigma}$, while the overall strength can be controlled by adjusting the thickness of a normal potential barrier between superconductor and lead, e.g.~by using  a few nm thick InGaAs layer \cite{PhysRevB.93.155402}.
Along these lines, we estimate that Zeeman fields of $B=1$\,meV produce couplings of $\Gamma_{\uparrow}=0.4$\,meV and $\Gamma_{\downarrow}=0$, giving rise to  $\gamma=\Delta$ and a length scale of $\ell_{\Gamma}\approx120$\,nm, see SM \cite{SM}. Thus, currently available heterostructures achieve both the necessary EP conditions and exhibit $a \ll \xi_{\Gamma}$,  assuring  that exceptional odd-$\omega$ pairs can homogeneously emerge in such systems.

\section{Conclusions}         
In conclusion, we have  shown the emergence of EPs in simple and physical NH superconducting systems based on conventional  superconductors. 
We have then demonstrated that such systems host odd-$\omega$ pairing purely due to the non-Hermiticity, which, at the EPs  becomes the only source of superconducting pairing, establishing the concept of exceptional odd-$\omega$ pairing. Finally, we showed how the spectral function can be used to detect both the emergence of EPs and measure exceptional odd-$\omega$ pairing.
Our work puts forward NH systems as a rich playground for generating odd-$\omega$ pairs, paving the way for NH engineering of dynamical superconducting states with enhanced and controlled properties.

\section{Acknowledgments}
We thank   E.~J.~Bergholtz, J.~C.~Budich, D.~Chakraborty, and L.~Potenciano and for insightful discussions.  We also acknowledge financial support from the Swedish Research Council (Vetenskapsr\aa det Grants No.~2018-03488 and No.~ 2021-04121) and the European Research Council (ERC) under the European Unions Horizon 2020 research and innovation programme (ERC-2017-StG-757553). 

\bibliography{biblio}
 \onecolumngrid

 

\section*{Supplementary material}

# Supplemental Material for "Exceptional odd-frequency pairing in conventional superconductors"

Jorge Cayao and Annica M. Black-Schaffer

*Department of Physics and Astronomy, Uppsala University, Box 516, S-751 20 Uppsala, Sweden*

In this supplementary material we provide details to support the results and conclusions of the main text. In particular, we give additional information on the derivation of the retarded self-energy.

## GREEN'S FUNCTIONS FOR AN OPEN SUPERCONDUCTOR SYSTEM

In the main text we utilize the superconducting pair amplitudes obtained following the Green's function approach [1, 2] for an open superconducting system. Our open system is modeled by an effective Hamiltonian that contains the Hamiltonian $H_S$ of an isolated conventional spin-singlet $s$-wave superconductor and a self-energy due to its coupling to a semi-infinite ferromagnetic lead, modeled by the Hamiltonian $H_L$. Geometrically, this open system consists of a two-dimensional (2D) junction along $z$: the closed system $H_S$ corresponds to a 2D superconductor, which can be thought of as having one site along $z$, while the ferromagnetic lead is semi-infinite along $z$ for negative $z$, see Fig. 1 in the main text.

The Green's function for our open system is given by [3]

$$G(\omega) = \left[\omega - H_S - V^\dagger g_L(\omega)V\right]^{-1},\tag{S1}$$

where $g_L(\omega) = (\omega - H_L)^{-1}$ is the Green's function of the semi-infinite lead and $V$ is the hopping matrix between the system and the lead (defined later). From Eq. (S1) we can define the effective Hamiltonian for the open superconductor system

$$H_{\rm eff} = H_S - \Sigma(\omega),\tag{S2}$$

where $H_S$ is expressed in Nambu space and given by

$$H_S = \xi_k \tau_z - \Delta \sigma_y \tau_y,\tag{S3}$$

while $\Sigma(\omega) = V^\dagger g_L(\omega)V$ is the spin-dependent self-energy in the superconductor due to the coupling to the ferromagnetic lead. Here, $\xi_k = \hbar^2 k^2/2m - \mu$ is the kinetic energy in the superconductor with $k = (k_x, k_y)$ and chemical potential $\mu$, $\sigma_i$ and $\tau_i$ are the Pauli matrices in spin and Nambu space, and $\Delta$ is the spin-singlet $s$-wave pair potential. We note that $H_S$ is written in the basis $\Psi_k = (c_{\uparrow,k}, c_{\downarrow,k}, c^\dagger_{\uparrow,-k}, c^\dagger_{\downarrow,-k})^{\rm T}$, where $c_{\sigma,k}$ annihilates an electronic state with spin $\sigma$ and momentum $k$.

To fully characterize the effective Hamiltonian $H_{\rm eff}$ in Eq. (S2) we need to find the self-energy, which in turn implies that we have to obtain the Green's function of the lead, $g_L$. But before obtaining these expressions we briefly discuss the matrix structure of the involved Green's functions in Nambu and spin spaces. The Green's function $G$ can, in principle, be obtained by performing the matrix inversion in Eq. (S1), which, due to the Nambu space, has the structure

$$G(\omega) = (\omega - H_{\rm eff})^{-1} = \begin{pmatrix} G_0 & F \\ \bar{F} & \bar{G}_0 \end{pmatrix},\tag{S4}$$

where $G_0$ and $\bar{G}_0$ represent the normal electron and hole components, while $F$ and $\bar{F}$ represent the electron-hole, or anomalous, components. Because the total open system includes an active spin degree of freedom, due to the ferromagnetic lead, the normal and anomalous components need to be matrices also in spin space, which in the chosen basis are given by

$$G_0(\omega) = \begin{pmatrix} [G_0]_{\uparrow\uparrow} & [G_0]_{\uparrow\downarrow} \\ [G_0]_{\downarrow\uparrow} & [\bar{G}_0]_{\downarrow\downarrow} \end{pmatrix}, \quad F(\omega) = \begin{pmatrix} F_{\uparrow\uparrow} & F_{\uparrow\downarrow} \\ F_{\downarrow\uparrow} & F_{\downarrow\downarrow} \end{pmatrix}.\tag{S5}$$

While $G_0$ determines the local density of states, $F$ directly gives the superconducting pair amplitudes, here with their spin structure explicitly expressed. This notation and knowledge is used in the main text when obtaining the

superconducting pair amplitudes. A similar discussion, as carried out in previous paragraph, applies for the Green's function of the lead $g_L$, which in Nambu space is given by

$$g_L(\omega) = (\omega - H_L)^{-1} = \begin{pmatrix} g_L^e & f_L \\ \bar{f}_L & g_L^h \end{pmatrix},$$ (S6)

where $g_L^{e(h)}$ represents the normal electron (hole) component, while $f$ and $\bar{f}$ represent the anomalous terms coding for the superconducting pair amplitudes in the lead. However, since the lead is not intrinsically superconducting, the pair amplitudes $f$ and $\bar{f}$ are zero and the $g_L$ is thus diagonal in Nambu space. We can therefore treat the electronic and hole blocks separately, which each have the spin structure

$$g_L^{e(h)}(\omega) = \begin{pmatrix} g_{\uparrow\uparrow}^{e(h)} & g_{\uparrow\downarrow}^{e(h)} \\ g_{\downarrow\uparrow}^{e(h)} & g_{\downarrow\downarrow}^{e(h)} \end{pmatrix}.$$ (S7)

To close this section, we have seen that the Green's functions are matrices in Nambu and spin spaces and they directly encode the superconducting pair amplitudes. To proceed, it is necessary to obtain the effective Hamiltonian in Eq. (S2) modeling the superconductor coupled to the ferromagnetic lead, which in turns requires calculation of the self-energy due to the lead. In order to find this self-energy we have to obtain the Green's function of the lead $g_L$, which we do in the next section.

## SELF-ENERGY DUE TO A FERROMAGNETIC SEMI-INFINITE LEAD

In this section we calculate the Green's functions of the lead $g_L^{e(h)}$. Because the lead is semi-infinite along the negative $z$-direction, it contains an infinite number of sites in this direction, with the Nambu Hamiltonian for each site $(i_L)$ given by the on-site $[H_L]_{i_L,i_L} = \xi_k^L \sigma_0 \tau_z + B \sigma_z \tau_z$, where $\xi_k^L = \hbar^2 k^2 / 2m - \mu_L$ is the kinetic term in the lead, with $k = (k_x, k_y)$, and chemical potential $\mu_L$. Further, $B$ represents the Zeeman energy, which could appear because the lead is ferromagnetic, due to contact with another ferromagnetic material, or by applying an external magnetic field. This thus implies that the Hamiltonian of the lead $H_L$ is an infinite matrix. Then, to find $g_L$ we proceed analytically, following a recursive approach as discussed e.g. in the Appendix of Ref. [4]. As already mentioned in the previous section, $g_L$ is diagonal in Nambu space and we can thus treat the electronic and hole blocks separately. For the electronic part we have

$$\begin{aligned} g_L^e &= (\omega - H_L^e)^{-1}, \\ &= (\omega - [H_L^e]_{1_L,1_L} - v^\dagger g_L^e v)^{-1}, \end{aligned}$$ (S8)

where $[H_L^e]_{1_L,1_L} = \xi_k^L \sigma_0 + B\sigma_z$ is the onsite energy at site $1_L$ of the lead, i.e. at the site in contact with $H_S$ (closest to the superconductor), and $v = -\sigma_0 t_z$ is the hopping matrix between sites along the $z$-direction in the lead. Note that we have here used the fact that the lead is semi-infinite and that is why $g_L^e$ also appears also on the right hand side. Then, by plugging Eq. (S7) into Eq. (S8) we get a system of equations whose solution fully determine $g_L^e$:

$$\begin{aligned} (\omega - \epsilon_\uparrow^e - t_z^2 g_{\uparrow\uparrow}^e) g_{\uparrow\uparrow}^e - t_z^2 g_{\uparrow\downarrow}^e g_{\downarrow\uparrow}^e &= 1, \\ (\omega - \epsilon_\uparrow^e - t_z^2 g_{\uparrow\uparrow}^e) g_{\uparrow\downarrow}^e - t_z^2 g_{\uparrow\downarrow}^e g_{\downarrow\downarrow}^e &= 0, \\ -t_z^2 g_{\downarrow\uparrow}^e g_{\uparrow\uparrow}^e + (\omega - \epsilon_\downarrow^e - t_z^2 g_{\downarrow\downarrow}^e) g_{\downarrow\uparrow}^e &= 0, \\ -t_z^2 g_{\downarrow\uparrow}^e g_{\uparrow\downarrow}^e + (\omega - \epsilon_\downarrow^e - t_z^2 g_{\downarrow\downarrow}^e) g_{\downarrow\downarrow}^e &= 1, \end{aligned}$$ (S9)

where $\epsilon_\uparrow^e = \xi_k^L + B$, $\epsilon_\downarrow^e = \xi_k^L - B$. A similar system of equations is obtained for the hole Green's function $g_L^h$, but then with onsite energies for holes, $\epsilon_\uparrow^e \to \epsilon_\uparrow^h = -\xi_k^L - B$ and $\epsilon_\downarrow^e \to \epsilon_\downarrow^h = -\xi_k^L + B$.

By solving the system of equations Eqs. (S9), we obtain the following retarded physical solutions

$$\begin{aligned} g_{\uparrow\uparrow}^{e(h)}(\omega) &= \frac{\omega - \epsilon_\uparrow^{e(h)} - \sqrt{(\omega - \epsilon_\uparrow^{e(h)})^2 - 4t_z^2}}{2t_z^2}, \\ g_{\downarrow\downarrow}^{e(h)}(\omega) &= \frac{\omega - \epsilon_\downarrow^{e(h)} - \sqrt{(\omega - \epsilon_\downarrow^{e(h)})^2 - 4t_z^2}}{2t_z^2}, \\ g_{\uparrow\downarrow}^{e(h)}(\omega) &= g_{\downarrow\uparrow}^{e(h)}(\omega) = 0, \end{aligned}$$ (S10)

for the elements of Green's function in the lead, $g_L$. As seen, only the diagonal, equal spin, entries are finite, which reflects the fact that spins in the lead are not coupled but only feel different Zeeman fields. It is worthwhile to point out that the expressions for the retarded Green's function $g_L$ have the following properties

$$g_{\sigma\sigma}^{e(h)}(\omega) = \begin{cases} \dfrac{1}{|t_z|}\left[\dfrac{\omega - \epsilon_\sigma^{e(h)}}{2|t_z|} - \mathrm{sgn}(\omega - \epsilon_\sigma^{e(h)})\sqrt{\left(\dfrac{\omega - \epsilon_\sigma^{e(h)}}{2|t_z|}\right)^2 - 1}\right], & |(\omega - \epsilon_\sigma^{e(h)})/2|t_z|| > 1 \\[4mm] \dfrac{1}{|t_z|}\left[\dfrac{\omega - \epsilon_\sigma^{e(h)}}{2|t_z|} - i\sqrt{1 - \left(\dfrac{\omega - \epsilon_\sigma^{e(h)}}{2|t_z|}\right)^2}\right], & |(\omega - \epsilon_\sigma^{e(h)})/2|t_z|| < 1, \end{cases} \tag{S11}$$

where $\sigma = \uparrow, \downarrow$. Before going further we notice that these forms of the Green's function have some consequences for the density of states, $\rho_L = -\frac{1}{\pi}\mathrm{Im}\,g_L$. In fact, for $|(\omega - \epsilon_\sigma^{e(h)})|/2|t_z| > 1$, i.e. first line in Eq. (S11), the Green's function $g_{\sigma\sigma}^{e(h)}$ exhibits fully real values and thus generates a vanishing local density of states. In contrast, for $|(\omega - \epsilon_\sigma^{e(h)})/2|t_z|| < 1$, within the bandwidth, i.e. second line in Eq. (S11), the Green's function $g_{\sigma\sigma}^{e(h)}$ develops an imaginary term, which produces a finite density of states equal to $[\rho_L(\omega)]_\sigma^{e(h)} = \dfrac{\theta(2|t_z| - |\omega - \epsilon_\sigma^{e(h)}|)}{|t_z|\pi}\sqrt{1 - \left(\dfrac{\omega - \epsilon_\sigma^{e(h)}}{2|t_z|}\right)^2}$.

Now we are in position to calculate the self-energy $\Sigma(\omega) = V^\dagger g_L(\omega)V$. Because $V$ is only finite between the nearest neighbor sites of the lead and the superconductor, it is possible to project the self-energy onto $H_S$, which can be done independently for the electron and hole parts independently,

$$\Sigma_{1_S 1_S}^{e(h)}(\omega) = \langle 1_S| V^\dagger |1_L\rangle \langle 1_L| g_L^{e(h)}(\omega) |1_L\rangle \langle 1_L| V |1_S\rangle , \tag{S12}$$

where $1_L$ denotes the first site of the lead, closest to the superconductor, while $1_S$ denotes the superconductor, which is only one site thick along the $z$-direction. We further set $\langle 1_L| V |1_S\rangle \equiv V_{1_L 1_S} = -t'\sigma_0$, where $t'$ is the hopping amplitude between sites $1_L$ in the lead and site $1_S$ in the superconductor. Equation (S12) implies that we only need the surface lead Green's function, the one at site $1_L$, which we extracted already in Eq. (S10). Using this we extract the self-energy as

$$\Sigma_{1_S 1_S}^{e(h)} = \begin{pmatrix} t'^2 g_{\uparrow\uparrow}^{e(h)} & 0 \\ 0 & t'^2 g_{\downarrow\downarrow}^{e(h)} \end{pmatrix}, \tag{S13}$$

where $g_{\uparrow\uparrow}^{e(h)}$ and $g_{\downarrow\downarrow}^{e(h)}$ are given by Eqs. (S10). By noticing that the Green's functions of the ferromagnetic lead develop real and imaginary terms, it is clear that the self-energy also develops real and imaginary terms. While both terms affect the Hamiltonian of the superconductor, only the imaginary term induces non-Hermitian (NH) physics. This imaginary self-energy contribution drastically alters the dynamics of quasiparticles in the superconductor, as it represents the rate at which quasiparticles in the superconductor escape through the lead. Finally, we note that the self-energy $\Sigma$ only has diagonal entries $\Sigma_{jj}^{e(h)}$ because of the particular fermionic Nambu basis we use. If we would instead have used a basis with the quasiparticles present after a Bogoliubov-Valatin transformation, which inherently mix electron- and hole-like properties, off-diagonal elements would have appeared in the self-energy.

In order to further investigate the role of the NH part of the self-energy, we perform some approximations for the Green's function of the lead, still maintaining its physical interpretation. We first take a version of the wide band limit, assuming $|(\omega - \epsilon_\sigma^{e(h)})/(2t_z)| \ll |(\mu_L - \sigma B)/(2t_z)| < 1$, which makes it possible to neglect the frequency and momentum dependence in the Green's function of the lead. Note, however, that we still assume $B$ and $\mu_L$ to be large enough to induce different imaginary terms in $g_L^{e(h)}$ for different spins. This approximation stems from the widely used wide-band approximation used in quantum transport [3, 5], which supports the assumptions we consider. Second, we point out that the real part of the self-energy is Hermitian and as such only introduces shifts into the superconductor Hamiltonian $H_S$. On the other hand, the imaginary part of the self-energy is NH and dramatically changes the $H_S$. Since we are primarily interested in investigating the NH properties of our open superconducting system, we focus only on the imaginary term of the self-energy and simply assume that the shifts introduced by the real part of the self-energy are already incorporated by an appropriate renormalization of $H_S$. Because we use tunable parameters to encode for the properties of $H_S$, such a renormalization is in fact already taken into account by sweeping over relevant parameter regimes.

From our discussion above and by using Eqs. (S13) and (S11), we can finally approximate the self-energy as

$$\Sigma \approx \begin{pmatrix} \Sigma^e(\omega = 0) & 0 \\ 0 & \Sigma^h(\omega = 0) \end{pmatrix}, \tag{S14}$$

where

$$\Sigma^{\mathrm{e,h}}(\omega = 0) = -i\Gamma\sigma_0 - i\gamma\sigma_z \,, \qquad\qquad\qquad (S15)$$

with $\Gamma = (\Gamma_\uparrow + \Gamma_\downarrow)/2$ and $\gamma = (\Gamma_\uparrow - \Gamma_\downarrow)/2$. Here we have denoted $\Gamma_\sigma = \pi|t'|^2\rho_{\mathrm{L}}^\sigma$, where $\rho_{\mathrm{L}}^{\uparrow(\downarrow)} = [1/(t_z\pi)]\sqrt{1 - [(\mu_{\mathrm{L}} \mp B)/(2t_z)]^2}$ is the spin-polarized surface density of states of the lead. As seen, the self-energy in Eq. (S14) contains negative imaginary terms, with its elements in the lower complex energy half-plane, and it thus represents a retarded self-energy evaluated at zero frequency, $\Sigma^r(\omega = 0)$. After obtaining this retarded self-energy in Eqs. (S14) and (S15), we are finally able to calculate the effective Hamiltonian for our open superconducting system using Eq. (S2). This, in turns, permits us to calculate the retarded Green's function by using Eq. (S1), whose off-diagonal components directly give the superconducting pair amplitudes. This retarded self-energy and the effective Hamiltonian are employed in the main text to discuss the impact of non-Hermiticity on both the generation of exceptional points (EPs) and odd-$\omega$ superconducting pairing. Here we note that deviations from the approximations used to arrive at Eq. (S14) are not expected to dramatically alter the results presented in the main text but might still provide some corrections such that e.g. the EPs appear in a renormalized parameter regime. We further note that non-Hermiticity due to disorder produces distinctly different NH self-energies [6, 7], and will therefore very likely be possible to distinguish from effects induced by the ferromagnetic lead.

## ESTIMATION OF COUPLINGS TO THE LEAD

In this part we discuss how to control the values of the couplings $\Gamma_{\uparrow,\downarrow}$, which appear in the diagonal entries of the self-energy in Eq. (S15) and are crucial for achieving the EP conditions presented in Eqs. (6) of the main text. Such control can be achieved by noting that these couplings are characterized by the hopping amplitude into the lead from the superconductor, $t'$, and by the spin-dependent surface density of states (DOS) of the ferromagnetic lead, $\rho_{\mathrm{L}}^{\uparrow,\downarrow}$. Because the lead is ferromagnetic, the DOS for up and down spins is distinct, allowing them to be controlled to a large degree by the strength of magnetic or Zeeman field. The chemical potential in the lead can additionally serve as a control parameter for surface DOS, which can be tuned by means of e.g. voltage gates. Furthermore, tunability of the hopping term $t'$ allows for control of the overall size of these couplings. It is important to here mention that, experimentally, $t'$ can straightforwardly be controlled by introducing a normal potential barrier between the lead and superconductor, with the barrier thickness directly, with exponential precision, tuning the strength of $t'$. We stress that this method has already been implemented in Ref. [8] for controlling the coupling between a superconductor and semiconductor, as well as other properties. Taken together these considerations supports the possibility to tune $\Gamma_{\uparrow,\downarrow}$ in experiments to achieve the EP conditions.

In order to visualize the necessary values of the couplings to reach the conditions for the EPs, in Fig. S1 we plot $\Gamma_{\uparrow,\downarrow}$ as a function of the Zeeman field $B$ and hopping parameter $t'$. The first observation in Fig. S1(a,b) is that $\Gamma_{\uparrow,\downarrow}$ exhibit distinct behavior as the Zeeman field $B$ increases. Another clear observation is that the size of the two couplings varies as the hopping $t'$ increases, as seen in Fig. S1 (a,b) and also in (c,d) where we fix $t'$. In particular, we find that for $t' = 1\,\mathrm{meV}$ and $B = 3\,\mathrm{meV}$, it is possible to obtain $\Gamma_\uparrow = 0.4\,\mathrm{meV}$ and $\Gamma_\downarrow = 0$, see cyan arrows in Fig. S1(d). These are exactly the values of the couplings required to reach the EP conditions discussed in the main text. The values of the couplings presented here are in range of the values reported in recent studies, see e.g. [8] that reports couplings of the order of $\approx 0.2 - 0.4\,\mathrm{meV}$ for superconductor-semiconductor hybrids, where the control of the couplings was carried out by introducing a normal barrier approximately $10\,\mathrm{nm}$ thick. This shows that achieving the conditions for the EP in Eq. (6) of the main text is clearly experimentally feasible.

Moreover, we point out that lower Zeeman fields can be used by adjusting the chemical potential in the lead $\mu_{\mathrm{L}}$, via e.g. voltage gates. This we show in Fig. S2, where we repeat the plots from Fig. S1 but now with a larger chemical potential in the lead $\mu_{\mathrm{L}}$. This fact is also evident by noting that the coupling depends on the expression for the surface DOS in the lead $\rho_{\mathrm{L}}^{\uparrow(\downarrow)}$. Again, this shows that achieving the conditions for the EP in Eq. (6) of the main text is clearly feasible in current experiments, especially since voltage gates are commonly used for transport measurements, see e.g. [8–11].

## NON-HERMITIAN LENGTH SCALE

As discussed in the main text, the inverse of the imaginary part of the eigenvalues in Eq. (5), i.e. the term determined by $\Gamma_{\uparrow,\downarrow}$, represent the lifetime or the average time a quasiparticle remains in the superconducting system before it

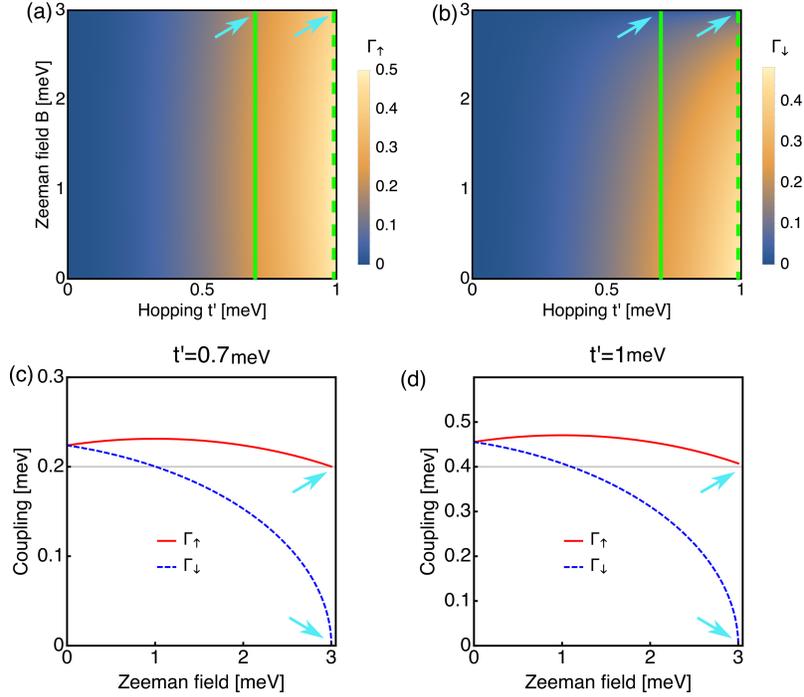

FIG. S1. Couplings $\Gamma_{\uparrow,\downarrow}$ as a function of Zeeman field $B$ and hopping parameter $t'$ (a,b) and as a function of Zeeman field for two different values of the hopping $t'$ (c,d). Vertical solid and dashed green lines in (a,b) mark the values of $t'$ used in (c,d), where we also find the couplings acquiring completely distinct behavior for $B = 3\,\mathrm{meV}$, with $\Gamma_{\uparrow} = 0.2\,\mathrm{meV}$ (grey line) and $\Gamma_{\downarrow} = 0$ in panels (a,c) and $\Gamma_{\uparrow} = 0.4\,\mathrm{meV}$ (grey line) and $\Gamma_{\downarrow} = 0$ in panels (b,d) with these points marked by cyan arrows. Parameters: $\mu_{\mathrm{L}} = 1\,\mathrm{meV}$, $t_z = 2\,\mathrm{meV}$.

escapes out into the lead. We can thus define a length scale $\ell_{\Gamma} = \hbar v_{\mathrm{F}}/\mathrm{Im}(E_n)$, with $v_{\mathrm{F}}$ the Fermi velocity in the normal state of the superconductor, associated with this non-Hermitian effect, telling us how far the non-Hermiticity penetrates the superconductor. In the EP regime, the imaginary part of the eigenvalues is set by $\Gamma = (\Gamma_{\uparrow} + \Gamma_{\downarrow})$, where $\Gamma_{\uparrow,\downarrow}$ are the diagonal, spin-dependent, entries of the non-Hermitian self-energy. Therefore, the length scale in the EP regime is given by $\ell_{\Gamma} = (\hbar v_{\mathrm{F}})/\Gamma$, where $v_{\mathrm{F}} = \hbar k_{\mathrm{F}}/m$, with $m$ the electron effective mass and $k_{\mathrm{F}}$ the Fermi wavevector determined by the Fermi energy, all in the superconductor. In order to induce a homogeneous, i.e. not depth dependent, non-Hermitian effect in the superconductor, the superconductor thickness $a$ has to be less than the non-Hermitian length scale, $a < \ell_{\Gamma}$. Thus, in order to induce a homogeneous non-Hermitian effect, it is desirable to have either very thin films or couplings that give large $\ell_{\Gamma}$. Along these lines, for InAs thin films with $m = 0.023\,m_e$, we get $v_{\mathrm{F}}$ at Fermi energies of $\mu_{\mathrm{S}} = 0.1\,\mathrm{meV}$, which, combined with $\Gamma_{\uparrow} = 0.4\,\mathrm{meV}$, and $\Gamma_{\downarrow} = 0$, allow us to estimate that this length scale is $\ell_{\Gamma} \approx 120\,\mathrm{nm}$. This is clearly much larger than the $a \sim 7\,\mathrm{nm}$ thickness of a typical InAs layer with proximity-induced superconductivity [8], thus guaranteeing that the whole superconducting layer feels a homogeneous non-Hermitian effect in these type of heterostructures. Stronger couplings produces shorter $\ell_{\Gamma}$, in line with the physical meaning of the non-Hermitian self-energy, where stronger couplings means shorter times for quasiparticles to spend in the superconducting system before they scape into the lead.

---

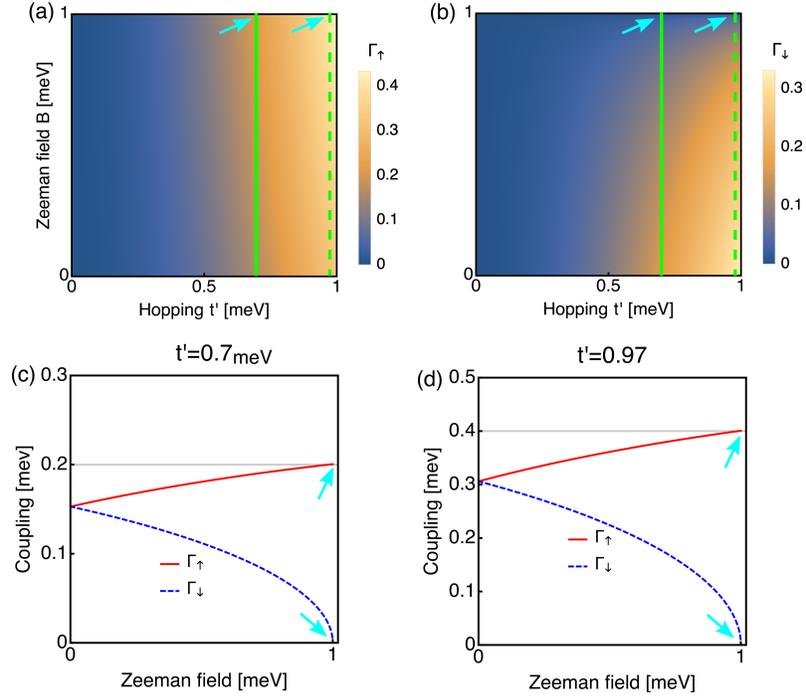

FIG. S2. Couplings $\Gamma_{\uparrow,\downarrow}$ as a function of Zeeman field $B$ and hopping parameter $t'$ (a,b) and as a function of Zeeman field for two different values of the hopping $t'$ (c,d). Vertical solid and dashed green lines in (a,b) mark the values of $t'$ used in (c,d), where we also find the couplings acquiring completely distinct behavior for $B = 1\,\mathrm{meV}$, with $\Gamma_{\uparrow} = 0.2\,\mathrm{meV}$ (grey line) and $\Gamma_{\downarrow} = 0$ in panels (a,c) and $\Gamma_{\uparrow} = 0.4\,\mathrm{meV}$ (grey line) and $\Gamma_{\downarrow} = 0$ in panels (b,d) with these points marked by cyan arrows. Parameters: $\mu_{\mathrm{L}} = 3\,\mathrm{meV}$, $t_z = 2\,\mathrm{meV}$.



\end{document}